\documentclass[10pt,twocolumn,letterpaper]{IEEEtran}
\usepackage[ left=1in,
                    right=1.01in,
                    top = 1 in,
                    bottom=1.07in]{geometry}
\usepackage{amsmath,epsfig}
\usepackage{algorithm}
\usepackage{bm}
\usepackage{tzplot}
\usepackage[usenames,dvipsnames]{pstricks}
\usepackage{epsfig}
\usepackage{pst-grad} 
\usepackage{pst-plot} 
\usepackage{subfigure}
\usepackage[capitalize]{cleveref}

\usepackage{subfigure}
\usepackage[T1]{fontenc}

\usepackage{graphics}
\include{psfig}
\usepackage{epsfig}
\usepackage{array}
\usepackage{psfrag}
\usepackage{amssymb}
\usepackage{color}
\usepackage{pstricks,pst-node,pst-text,pst-3d,pst-plot}
\usepackage{cite}
\usepackage{ulem}
\definecolor{dgreen}{rgb}{0, 0.8, 0.1}

\usepackage{amsmath}
\usepackage{breqn}

\begin{document}

\title{Target Detection in Sea Clutter with Application to Spaceborne SAR Imaging \\
\author{\IEEEauthorblockN{Shahrokh Hamidi}\\
\IEEEauthorblockA{\textit{Department of Electrical and Computer Engineering, University of Waterloo}\\
\textit{Waterloo, Ontario, Canada} \\
shahrokh.hamidi@uwaterloo.ca}
}
}

\maketitle
\thispagestyle{empty}

\begin{tikzpicture}[remember picture, overlay]
      \node[font=\small] at ([yshift=-1cm]current page.north)  {This paper has been published in the 2025 International Conference on Radar, Antenna, Microwave, Electronics, and Telecommunications (ICRAMET). \copyright IEEE};
\end{tikzpicture}
\begin{abstract}
In this paper, the challenging task of target detection in sea clutter is addressed. We analyze the statistical properties of the signals which have been received from the scene and based on that, we model the amplitude of the signals that have been reflected from the background sea clutter according to several well-known probability distribution functions.
Next, by exploiting the Kullback-Leibler (KL) divergence metric as a goodness-of-fit test, we will demonstrate that among the proposed probability distributions, the Weibull distribution can model the statistical properties of the background sea clutter with higher accuracy.
Subsequently, we utilize the aforementioned information to design an adaptive threshold based on the Constant False Alarm Rate (CFAR) algorithm to detect the energy of the targets which have been buried in the sea clutter.
Thorough analysis of the experimental data gathered from the Canadian RADARSAT-1 satellite demonstrates the overall effectiveness of the proposed method. 
\end{abstract}


\section{Introduction}
As an all-weather, day-night, and active device, Synthetic Aperture Radar (SAR) is capable of providing crucial information from the surface of the Earth \cite{Cumming}. Following the image reconstruction, post processing procedure is performed to obtain more critical information from the scene. Target detection is one of the areas that requires further analysis of the reconstructed images. In this paper, we specifically focus on the targets buried in sea clutter. The goal is to establish a mechanism to extract the energy of the targets from sea clutter. Several techniques have been reported in the literature, such as the wavelet-based analysis \cite{Wavelet} and fixed threshold-based method \cite{fixed_threshold}. The problem with using a fixed threshold to detect targets from clutter is that some of the key features of the desired targets can be ignored while the undesired signals are retained. Moreover, methods such as wavelet-based analysis are not able to capture the statistical properties of the signals. This, in turn, hinders their ability to obtain all the critical information about the desired targets. 

In this paper, we utilize an approach based on the Constant False Alarm Rate (CFAR) method \cite{Mahafza} which creates an adaptive threshold to extract the energy of the desired targets while suppressing the energy of the undesired signals. 
However, in order to be able to implement the CFAR-based adaptive threshold, we need to obtain the information about the statistical properties of the undesired signals which in this case is the sea clutter. To tackle this issue, we study several well-known probability density functions, namely, Weibull, Log-normal,  Inverse Gaussian, Gamma, and Rayleigh and demonstrate that the Weibull distribution is a more appropriate distribution to model the statistical properties of the sea clutter.
In \cite{SAR_Clutter, CFAR_Weibull, Shahrokh_MSTAR, ground_clutter}, the authors have studied the statistical properties of the reflected SAR signals. In this paper, however, we specifically focus on sea clutter and also we present more statistical models which distinguishes our work from \cite{SAR_Clutter, CFAR_Weibull, Shahrokh_MSTAR}. 
We further describe the SAR imaging procedure in detail and present the experimental results to verify the effectiveness and accuracy of the proposed models which makes the material presented in this paper distinct from \cite{SAR_Clutter, CFAR_Weibull, Shahrokh_MSTAR, ground_clutter}.
In contrast to \cite{ground_clutter}, we will not consider the K distribution. The reason for this is that, the K distribution is formed by compounding two separate probability distributions, one representing the radar cross-section, and the other representing the speckle noise. However, through experimental results, we demonstrate that the Weibull distribution can model the sea clutter highly accurately. As a result, the K distribution is no longer considered as a possible candidate to model the statistical properties of the sea clutter in this paper. In fact, after removing the effect of the speckle noise the K distribution is no longer capable of modeling the sea clutter accurately. 

Several statistical models are presented to describe the background sea clutter and at the end we demonstrate that the Weibull distribution is the most appropriate distribution to model the statistical properties of the background sea clutter. In order to determine which probability distribution can model the background sea clutter with higher accuracy, we exploit the Kullback Leibler (KL) divergence metric as a goodness-of-fit test. 
We then utilize the Weibull distribution and establish an adaptive threshold to extract the energy of the desired targets from the background sea clutter.
The verification of the proposed approach is based on the real data gathered from the Canadian RADARSAT-1 SAR satellite \cite{RADARSAT}. The data has been gathered from English Bay in Vancouver Canada in which several ships have been located inside water which creates a perfect scenario to evaluate the results.

The organization of the paper is as follows. 
In Section \ref{Model Description}, we present the system model as well as the Range-Doppler algorithm to be used to reconstruct the SAR images from the raw data. In Section \ref{Statistical Modeling}, we address the statistical analysis of the SAR data. Finally, Section \ref{experimental results} has been dedicated to the experimental results based on the real data gathered from the RADARSAT-1 SAR satellite which is followed by the concluding remarks.

\section{Model Description And Image Formation}\label{Model Description}
The signal transmitted by the radar is a chirp signal which after being reflected from a point target and upon down-conversion is described as 
\begin{dmath}
\label{RX_signal}
s(t, \eta) = \sigma \; w_r(t - \frac{2R(\eta)}{c})w_a(\eta - \eta_c)\times
e^{\displaystyle  -j 4 \pi f_c\frac{R(\eta)}{c} + j \pi \beta t^2 + j 4 \pi \beta t \frac{R(\eta)}{c}},
\end{dmath}
where $f_c$ is the carrier frequency and the parameter $\beta$ is given as $b/T$, in which $b$ and $T$ stand for the bandwidth and the chirp time, respectively. In addition, $w_r$ is a rectangular window with length $T$ and $t$ is referred to as fast-time parameter.
In addition, $\sigma$ is the complex radar cross section for the point reflector, $\eta$ is referred to as slow-time parameter and $R(\eta)$ is the instantaneous radial distance between the radar and the target which is given as  $R(\eta) = \sqrt{R^2_0 + v^2(\eta - \eta_c)^2}$. Moreover, $w_a$ is a rectangular window with its length equal to the synthetic aperture length divided by $v$, where $v$ is the speed of the satellite.
The Range-Doppler algorithm is the first algorithm that was developed for spaceborne SAR image reconstruction \cite{Cumming_RD}.
The first step in the Range-Doppler algorithm is to perform range compression which results in \cite{Cumming}
\begin{dmath}
\label{Range_Compressed}
s_{\rm rc}(t, \eta) = \sigma \; p_r(t - \frac{2R(\eta)}{c})w_a(\eta - \eta_c) 
           e^{\displaystyle  -j 4 \pi f_c \frac{R(\eta)}{c}},
\end{dmath}
where $p_r(x) = \frac{sin(\pi x)}{\pi x}$.
Next, we should compensate for the Range Cell Migration (RCM) phenomenon which is the result of coupling between the range and azimuth directions that is created because of the non-zero squint angle for the antenna. Consequently, following the RCM compensation, the signal in the Doppler domain is described as \cite{Cumming}
\begin{dmath}
\label{RCMC}
S_{\rm rcmc}(t, f_\eta) = \sigma \; p_r(t - \frac{2R_0}{c})W_a(f_{\eta} - f_{\eta_c}) \times 
           e^{\displaystyle -j \frac{4 \pi f_c R_0}{c}} e^{\displaystyle  j \pi \frac{f^2_{\eta}}{K_a}},
\end{dmath}
in which $K_a = \frac{2v^2}{\lambda R_0}$.
The final step for image formation in the Range-Doppler algorithm, is to compress the data in the azimuth direction which upon performing this stage the compressed signal in the range and azimuth directions is given as \cite{Cumming}
\begin{dmath}
\label{Image_RD}
s_{\rm rcac}(t, \eta)  = \sigma \; p_r(t - \frac{2R_0}{c})p_a(\eta) e^{\displaystyle  -j \frac{4 \pi f_c R_0}{c}} \times 
          e^{\displaystyle  j 2 \pi f_{\eta_c} \eta},
\end{dmath}
where $f_{\eta_c}$ is the Doppler centroid frequency \cite{Cumming}.

\section{Statistical Modeling}\label{Statistical Modeling}
In this section, we attempt to model the amplitude of the reflected signal from sea clutter. The main candidate is the Weibull distribution which  its probability  density functions is described as   \cite{Distribution} 
\begin{dmath}
\label{Weibull}
p_W({x}| \alpha, \beta) =  \displaystyle \frac{\alpha}{\beta} {\left(\frac{x}{\beta}\right)}^{\alpha-1}e^{ \displaystyle - \left(\frac{x}{\beta}\right)^{\alpha}},  \;\;\; x \ge 0, \;\;\alpha, \beta > 0.
\end{dmath}
The other possible options, for consideration to model the sea clutter, are the Log-normal, Inverse Gaussian,  Gamma,  and  Rayleigh distributions with the following probability  density functions  \cite{Distribution}
\begin{dmath}
\label{lognormal}
p_{\rm LN}({x}| \eta, \gamma) =  \displaystyle \frac{1}{x\eta \sqrt{2 \pi}} e^{ \displaystyle - \frac{(\log x -\gamma)^2}{2 \eta^2}},  \;\;\; x > 0, \;\; \eta > 0.
\end{dmath}
\begin{dmath}
\label{InverseGaussian}
p_{\rm IG}({x}| \mu, \lambda) =  \displaystyle \sqrt{\frac{\lambda}{2 \pi x^3}} e^{ \displaystyle - \frac{\lambda(x -\mu)^2}{2 \mu^2 x}},  \;\;\; x > 0, \;\; \mu, \lambda > 0.
\end{dmath}
\begin{dmath}
\label{Gamma}
p_G(x| a, b) = \frac{1}{b^a \Gamma(a)} x^{a-1}  e^{-x/b}, \;\;\; x>0, \;\; a, b > 0.
\end{dmath}
\begin{dmath}
\label{Rayleigh}
p_R(x| \sigma) = (x/ \sigma^2) e^{\displaystyle -\frac{x^2}{2\sigma^2}}, \;\;\; x \ge 0, \;\; \sigma > 0.
\end{dmath}
Based on the given probability of false alarm ($p_{\rm fa}$), we can calculate the adaptive threshold $T_a$ as  \cite{Mahafza} 
\begin{eqnarray}
\label{pfa}
p_{\rm fa} = p(X > T_a| H_0) = \int_{ T_a}^{\infty}  p(X| H_0) \; dX,
\end{eqnarray}
where $X$ describes the statistics of the cell under test and $T_a$ is the adaptive threshold. Furthermore, $H_0$ represents the null hypothesis. 
By using the experimental data that we present in this paper, we will demonstrate that among the proposed distributions, the Weibull distribution is the best distribution to model the sea clutter. Therefore, the calculation of the adaptive threshold is only performed for the Weibull distribution. 
Consequently, based on (\ref{pfa}), the adaptive threshold $T_{\rm aW}$ for the Weibull distribution, which has been given in (\ref{Weibull}), is calculated as
\begin{eqnarray}
\label{thr_weibull}
T_{\rm aW} = \beta [ \log \left(\frac{1}{p_{\rm fa}}\right)]^{\displaystyle \frac{1}{\alpha}}.
\end{eqnarray}
The maximum likelihood estimates for the parameters of the Weibull distribution, given in (\ref{Weibull}), are expressed as \cite{Weibull_ML}
\begin{eqnarray}
\label{params_weibull}
\hat{\beta} = \left(\frac{1}{N} \sum_{i = 1}^{N} x^{\hat{\alpha}}_i\right)^{\displaystyle \frac{1}{\hat{\alpha}}}, \nonumber \\
\hat{\alpha} = \displaystyle \frac{N}{\frac{1}{\hat{\beta}} \sum_{i = 1}^{N} x^{\hat{\alpha}}_i \log x_i - \sum_{i = 1}^{N} \log x_i},
\end{eqnarray}
in which $x_i$ is the $i^{\rm th}$ realization of the random variable $x$.  
In this paper, we implement the 2D CFAR \cite{Mahafza} algorithm.  
Subsequently, the decision making process for the cell under test is expressed as \cite{CFAR_Weibull, Shahrokh_MSTAR}
\begin{equation}
\label{detection}
 X_{\rm CUT}  \mathop{\lessgtr}_{H_1}^{H_0}  \mu_c + \sigma_c  Q,
\end{equation}
where $X_{\rm CUT}$ is  the amplitude of the cell under test. Furthermore, $\mu_c$ and $\sigma_c$ are the sample mean and standard deviation computed from the clutter data of the local background. 
Moreover, $Q = \frac{T_{\rm aW}}{\hat{\mu}}$ is the detector design parameter which defines the $p_{\rm fa}$ and is set empirically in which $ T_{\rm aW}$ is the adaptive threshold which has been described in (\ref{thr_weibull}) and $\hat{\mu}$ denotes the mean value estimated from the underlying clutter model. In addition, the $H_0$ and $H_1$ are the null and alternative hypotheses, respectively \cite{Kay, Mahafza}.    
The null hypothesis, $H_0$, represents the case in which the cell under test has been occupied by clutter only and the alternative hypothesis, $H_1$, on the other hand, is for the case in which the target is present and the cell under test contains the energy of the target.   
In the next section, we present the experimental results. 
\section{Experimental Results}\label{experimental results}
\begin{figure}
\centering
\begin{tikzpicture}[yshift=0.00001cm][font=\large]
\node(img1) {\includegraphics[height=3.5cm,width=5.5cm]{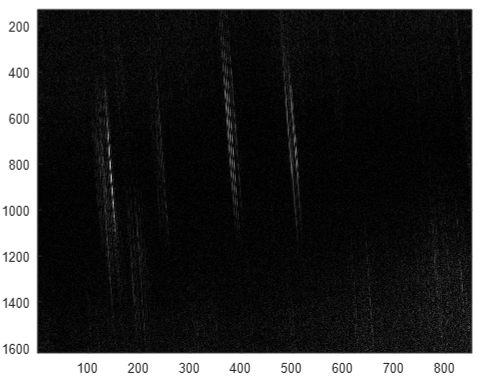}};
  \node[left=of img1, node distance=0cm, rotate = 90, xshift=2.2cm, yshift=-0.8cm,font=\color{black}] {{Along-Track [Samples]}};
  \node[below=of img1, node distance=0cm, xshift=0cm, yshift=1.1cm,font=\color{black}] {{Slant-Range [Samples]}};
\end{tikzpicture}
\caption{The  range compressed data for several strong isolated targets. In this image, the isolated targets are several ships in English Bay.}
\label{fig:Range_Compressed}
\end{figure}

\begin{figure}
\centering
\begin{tikzpicture}[yshift=0.00001cm][font=\large]
  \node (img1)  {\includegraphics[height=4cm,width=6cm]{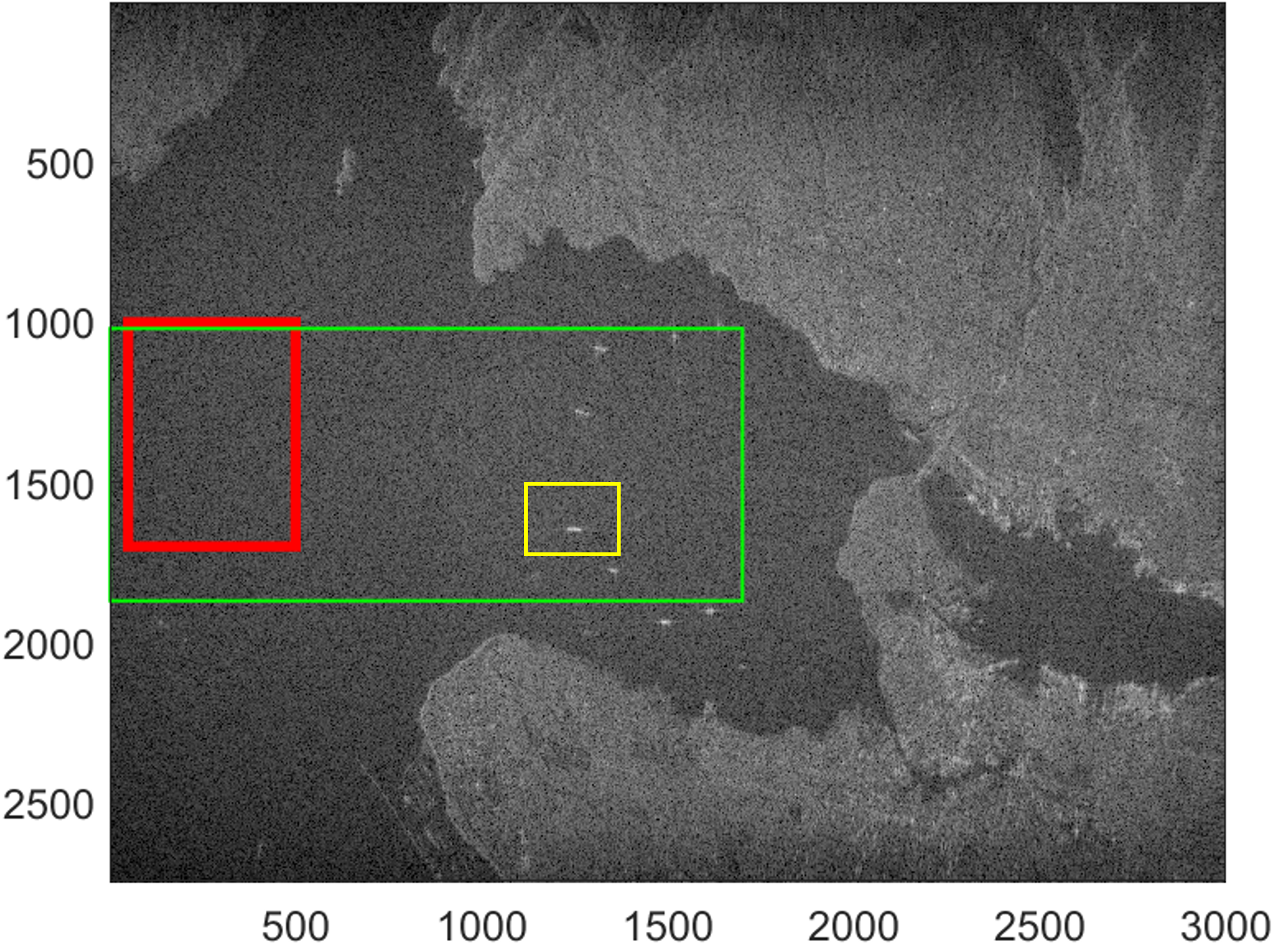}};
  \node[left=of img1, node distance=0cm, rotate = 90, xshift=2.2cm, yshift=-0.8cm,font=\color{black}] {{Along-Track [Samples]}};
  \node[below=of img1, node distance=0cm, xshift=0cm, yshift=1.1cm,font=\color{black}] {{Slant-Range [Samples]}};
\end{tikzpicture}
\caption{The reconstructed image from English Bay, in Vancouver Canada, based on the Range-Doppler algorithm.}
\label{fig:EnglishBay_Speckle}
\end{figure}

\begin{figure}
\centering
\begin{tikzpicture}[yshift=0.00001cm][font=\large]
  \node (img1)  {\includegraphics[height=5cm,width=7cm]{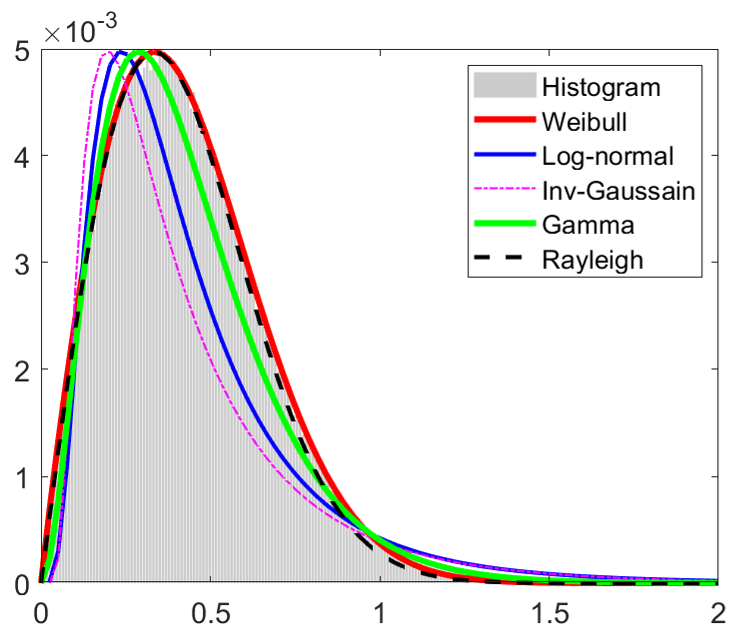}};
  \node[left=of img1, node distance=0cm, rotate = 90, xshift=1.1cm, yshift=-0.8cm,font=\color{black}] {{Frequency}};
  \node[below=of img1, node distance=0cm, xshift=0cm, yshift=1cm,font=\color{black}] {{$\sqrt{{I}^2 + {Q}^2}$}};
\end{tikzpicture}
\caption{The histogram of the experimental data inside the red box which is related to the sea clutter as well as the estimated probability density functions based on (\ref{Weibull}), (\ref{lognormal}), (\ref{InverseGaussian}), (\ref{Gamma}), and (\ref{Rayleigh}). The $I$ and $Q$ represent the imaginary and quadratic part of the estimated value for the complex radar cross section. }
\label{fig:hist}
\end{figure}

In this section, we present the result of SAR image reconstruction based on the experimental data gathered from the Canadian RADARSAT-1 satellite. The data is from Vancouver Canada. 
The specifications of the RADARSAT-1 satellite are given in Table.\ref{Tab:Table}.
\begin{table}
  \centering
  \caption{RADARSAT-1's Parameters}\label{Tab:Table}
  \begin{center}
    \begin{tabular}{| l | l | l |}

    \hline
    Parameters &  & Values  \\ \hline
    Center frequency ($\rm GHz$) & $\rm f_c$ & $5.3$ \\ \hline
    Radar sampling rate ($\rm MHz$) & $\rm Fr$ & $32.317$  \\ \hline
    Pulse repetition frequency ($\rm Hz$) & $\rm PRF$ & $1256.98$ \\ \hline
    Slant range of first radar sample ($\rm km$) & $R_0$ & $988.65$  \\ \hline
    FM rate of radar pulse (${\rm MHz}/\mu s$) & $ \beta$     &  $0.72135$   \\ \hline
    Chirp duration ($ \mu s$)  &  $\rm T$  &  $41.75$   \\ \hline
    Satellite velocity ($\rm m/s$) &   $\rm v$  &  $7062$    \\ \hline
    Bandwidth ($\rm MHz$)  &  $\rm b$   & $30.116$ \\ \hline
    \hline
    \end{tabular}
\end{center}
\end{table}
We select the data related to English Bay in Vancouver, Canada. The reason for this choice is that, there are several ships in this scene that can play the role of isolated strong reflectors which will assist with  observing the effect of the RCM clearly and will also be utilized later to perform the Doppler centroid frequency estimation. 
Furthermore, the ships in the sea water are the main subject of the paper. In other words, we attempt to detect the ships in sea clutter which is the main goal of the paper.
In order to reconstruct the image, we apply the Range-Doppler algorithm to the raw data. Fig.~\ref{fig:Range_Compressed}-(a) shows the range compressed data based on (\ref{Range_Compressed}).
The range compressed energy of the ships can be seen as several skewed vertical lines. The skew demonstrates the effect of the RCM.
In order to perform the RCM compensation we need to estimate the unambiguous Doppler centroid frequency.
When the Doppler centroid frequency is larger than the PRF of the RADAR, there will be an ambiguity in the Doppler centroid frequency estimation.
One way to estimate the unambiguous value for the Doppler centroid frequency is by analyzing the trajectory of the strong isolated targets.
Fig.~\ref{fig:Range_Compressed} shows the range compressed image of several ships in sea water.
The skew in the trajectory of these targets is due to the non-zero Doppler centroid frequency. The energy of each of these targets is spread over several different range cells.
The slope can easily be calculated and is equal to $0.034$ range samples per azimuth samples. To estimate the Doppler centroid frequency, the first step is to multiply the slope by $\rm \frac{c}{2Fr}$ to convert the slope from range samples to range distance and then multiply it by $\rm PRF$ to convert it from azimuth samples to azimuth time.
Hence, we have $\frac{dR(\eta)}{d\eta} = 198.23 \; \rm m/s$. As a result, from $f_{\rm dc} = -\frac{2}{\lambda}\frac{dR(\eta)}{d\eta}$ \cite{Cumming}, the Doppler centroid frequency can be calculated as $f_{\rm dc} = -7009 \; \rm Hz$.
\begin{figure}
\centering
\begin{tikzpicture}[yshift=0.00001cm][font=\large]
\node(img2) {\includegraphics[height=3cm,width=5cm]{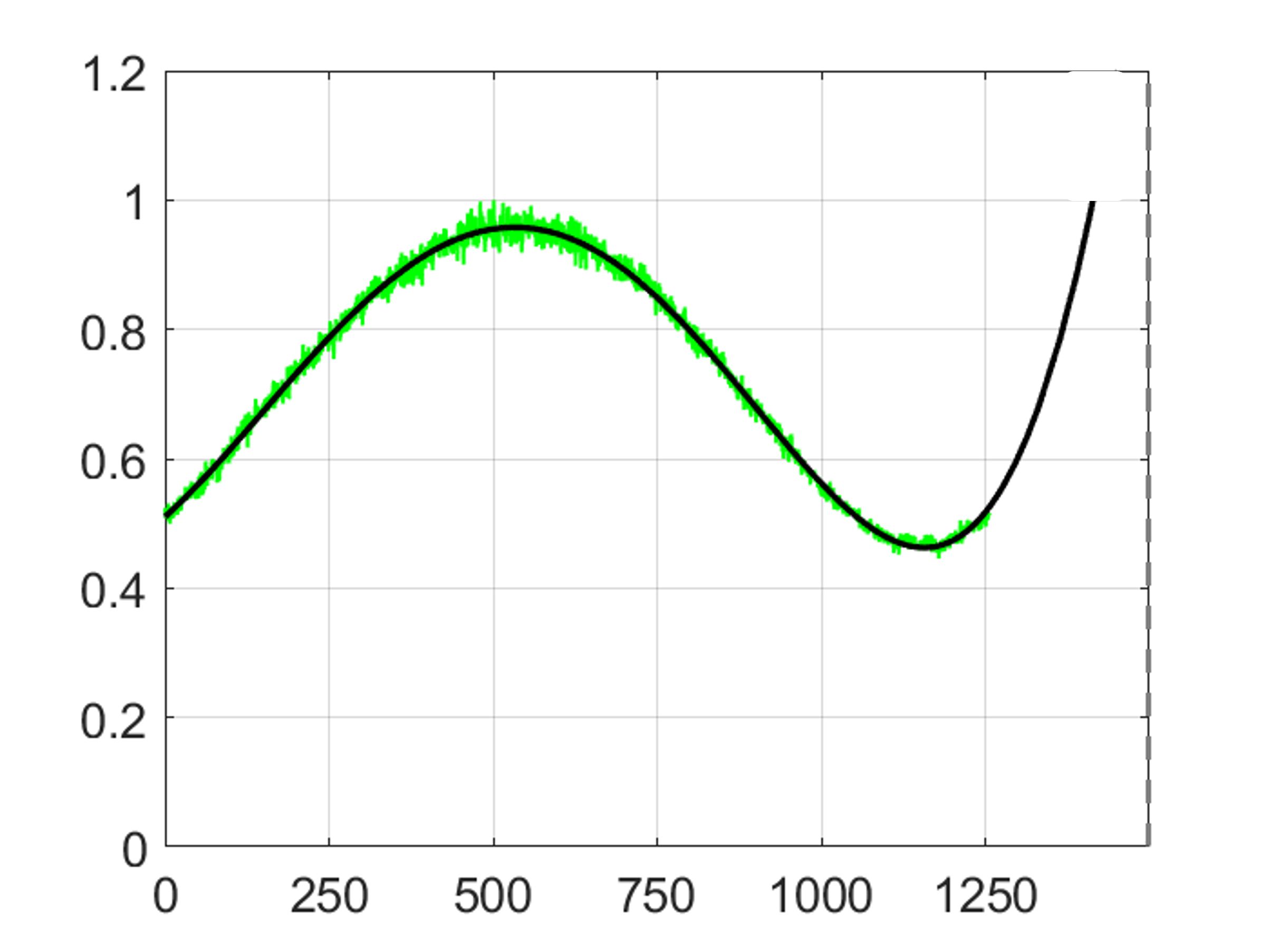}};
  \node[left=of img1, node distance=0cm, rotate = 90, xshift=1.9cm, yshift=-1.1cm,font=\color{black}] {{ Normalized Power}};
  \node[below=of img1, node distance=0cm, xshift=0cm, yshift=1.1cm,font=\color{black}] {{Frequency [Hz]}};
\end{tikzpicture}
\caption{The amplitude-based estimation of the fractional part of the Doppler centroid frequency with its maximum value at $f^{\prime}_{\rm dc} = 531 \; \rm Hz$.}
\label{fig:fdc}
\end{figure}
Next, we estimate the fractional part of the Doppler centroid frequency which is $f^{\prime}_{\rm dc}$ \cite{Cumming}. The fractional part is essential in focusing the energy of the targets in the azimuth direction. 
Fig.~\ref{fig:fdc} illustrates the power spectrum of the data in azimuth direction versus slow-time frequency component $f_{\eta}$. In order to reduce the effect of the noise, we have performed an averaging over the power spectrum of $2048$ range cells. In order to be able to calculate the frequency component at which the signal reaches its maximum value, we have performed curve fitting.
From Fig.~\ref{fig:fdc}, we can estimate the fractional part of the Doppler centroid frequency as $f^{\prime}_{\rm dc} = 531\; \rm Hz$.
Finally, at the last stage, we perform the azimuth localization based on (\ref{Image_RD}). Consequently, the reconstructed image based on the Range-Doppler algorithm is obtained which has been presented in Fig.~\ref{fig:EnglishBay_Speckle}.
However, the image presented in Fig.~\ref{fig:EnglishBay_Speckle} suffers from speckle noise \cite{Cumming, Shahrokh_Speckle_1, Shahrokh_Speckle_2}. In order to remove the effect of the speckle noise, we introduce a 2D $m \times n$ filter and slide it over the reconstructed image while solving the following optimization problem,
\begin{eqnarray}
\label{speckle}
\!\min_{a}        &\qquad&    \sum_{i=1}^{n} \sum_{j=1}^{m}| a_{\rm ij}-a|,
\end{eqnarray}
where $a_{\rm ij}$ is the value for the $(\rm ij)^{\rm th}$ pixel and $a$ is the value selected by the optimization problem for the $(\lfloor \frac{n-1}{2} \rfloor + 1, \; \lfloor \frac{m-1}{2} \rfloor + 1)^{\rm th}$ pixel. In fact, by solving the optimization problem in (\ref{speckle}) we apply a 2D median filter to the image. In other words, we replace the value of each pixel with the median of the neighboring pixels. This is proved to be a powerful method to decrease the effect of the speckle noise which, as a result, the fine structures of the image can be revealed. 
It should be noted that, the common practice to alleviate the effect of the speckle noise in SAR images is multi-look processing \cite{Cumming}. However, multi-look processing sacrifices the azimuth resolution in order to remove the effect of the speckle noise. In contrast, the median filtering approach, which we have proposed, can efficiently remove the speckle noise while it leaves the azimuth resolution intact. Furthermore, the median filtering does not smear the edges in the image which is considered to be a significant property. 
\begin{figure}
\centering
\begin{tikzpicture}[yshift=0.00001cm][font=\large]
  \node (img1)  {\includegraphics[height=3cm,width=5cm]{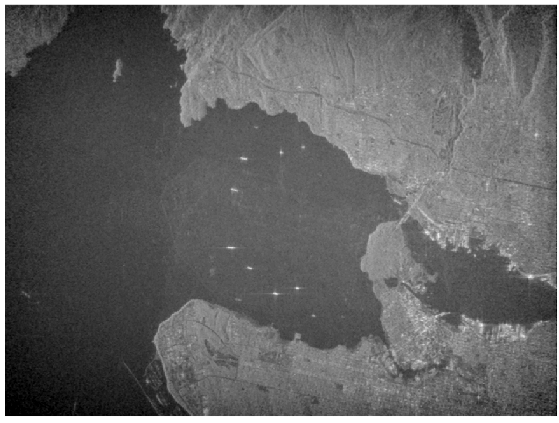}};
  \node[left=of img1, node distance=0cm, rotate = 90, xshift=1.3cm, yshift=-0.8cm,font=\color{black}] {{Along-Track}};
  \node[below=of img1, node distance=0cm, xshift=0cm, yshift=1.1cm,font=\color{black}] {{Slant-Range}};
\end{tikzpicture}
\caption{The result of the speckle noise reduction process, based on (\ref{speckle}), for the reconstructed image of English Bay, in Vancouver Canada, shown in Fig.~\ref{fig:EnglishBay_Speckle}.}
\label{fig:EnglishBay}
\end{figure}
Fig.~\ref{fig:EnglishBay} shows the result of speckle noise reduction for the reconstructed image depicted in Fig.~\ref{fig:EnglishBay_Speckle}.  To remove the effect of the speckle noise, we have applied the filter, given in (\ref{speckle}), with $m=n=6$.
In order to analyze the effect of the sea clutter on target detection, we focus on the specific part of the  reconstructed image shown in Fig.~\ref{fig:EnglishBay_Speckle} which contains several ships in sea water. The related part of the data-set has been highlighted as the area inside the green rectangle in  Fig.~\ref{fig:EnglishBay_Speckle}. The area inside the red rectangle contains the energy from sea water and is utilized to analyze the statistical properties of the sea clutter. Fig.~\ref{fig:hist} shows the histogram for part of the image related to the area which has been located inside the red rectangle. The estimated probability density functions for the Weibull,  Log-normal, Inverse Gaussian, Gamma, and Rayleigh distributions are based on (\ref{Weibull}), (\ref{lognormal}), (\ref{InverseGaussian}), (\ref{Gamma}), and (\ref{Rayleigh}), respectively.
The estimated parameters for the given probability density functions are presented in Table.\ref{Tab:Table_hist}.
\begin{table}
  \centering
  \caption{The estimated parameters for the given probability density functions}\label{Tab:Table_hist}
  \begin{center}
    \begin{tabular}{| l | l | l |}

    \hline
    Distribution &  &   \\ \hline
    Weibull &  $\alpha = 1.9521$       &        $\beta = 0.4835$ \\ \hline
    Log-normal & $\gamma =  -1.0201$    &       $\eta  = 0.6484$  \\ \hline
    Inv-Gaussian & $\mu =   0.4286 $    &      $\lambda =  0.7422$ \\ \hline
    Gamma & $a = 3.0486 $ & $b =  0.1406$  \\ \hline
    Rayleigh & $\sigma = 0.3337 $     &    \\ \hline
    \hline
    \end{tabular}
\end{center}
\end{table}
In order to understand which distribution function is the most appropriate option to model the statistical properties of the sea clutter, we exploit the KL distance as a measure of fitness. For the estimated and empirical probability density functions described as $p_e(x)$ and $p_d(x)$, respectively, the KL distance is expressed as 
\begin{equation}
\label{KL}
D_{\rm KL}{(p_d(x) || p_e(x))} = \int_{- \infty}^{\infty}   p_d(x) \log \left( \frac{p_d(x)}{p_e(x)} \right)   dx.
\end{equation}
The values for the KL distance for the given probability density functions have been calculated based on (\ref{KL}) and have been presented in Table.\ref{Tab:Table_KL}.
\begin{table}
  \centering
  \caption{The values of the KL distance for the given  probability density functions}\label{Tab:Table_KL}
  \begin{center}
    \begin{tabular}{| l | l | l |}

    \hline
    Distribution &  KL Distance  \\ \hline
    Weibull &  $0.0017$        \\ \hline
    Log-normal & $0.0270$     \\ \hline
    Inv-Gaussian & $0.0406$     \\ \hline
    Gamma & $0.0089 $  \\ \hline
    Rayleigh & $0.0029$     \\ \hline
    \hline
    \end{tabular}
\end{center}
\end{table}
From Table.\ref{Tab:Table_KL}, it is evident that compared to the other probability distributions, the Weibull distribution, with the lowest KL distance, can model the statistical properties of the sea clutter with higher accuracy. Therefore, we only consider the Weibull distribution for the CFAR-based analysis of target detection in sea clutter. 
Fig.~\ref{fig:cfar_ships} illustrates the result of applying the 2D CFAR algorithm to the part of the data-set inside the green rectangle in Fig.~\ref{fig:EnglishBay_Speckle}. The result presented in Fig.~\ref{fig:cfar_ships}-(b) is remarkable in a sense that, by modeling the sea clutter based on the Weibull distribution and by  exploiting the 2D CFAR technique we have been able to detect the ships in sea clutter with high accuracy.  
\begin{figure}
\begin{tikzpicture}[yshift=0.00001cm][font=\small]
  \node (img1)  {\includegraphics[height=3cm,width=3.5cm]{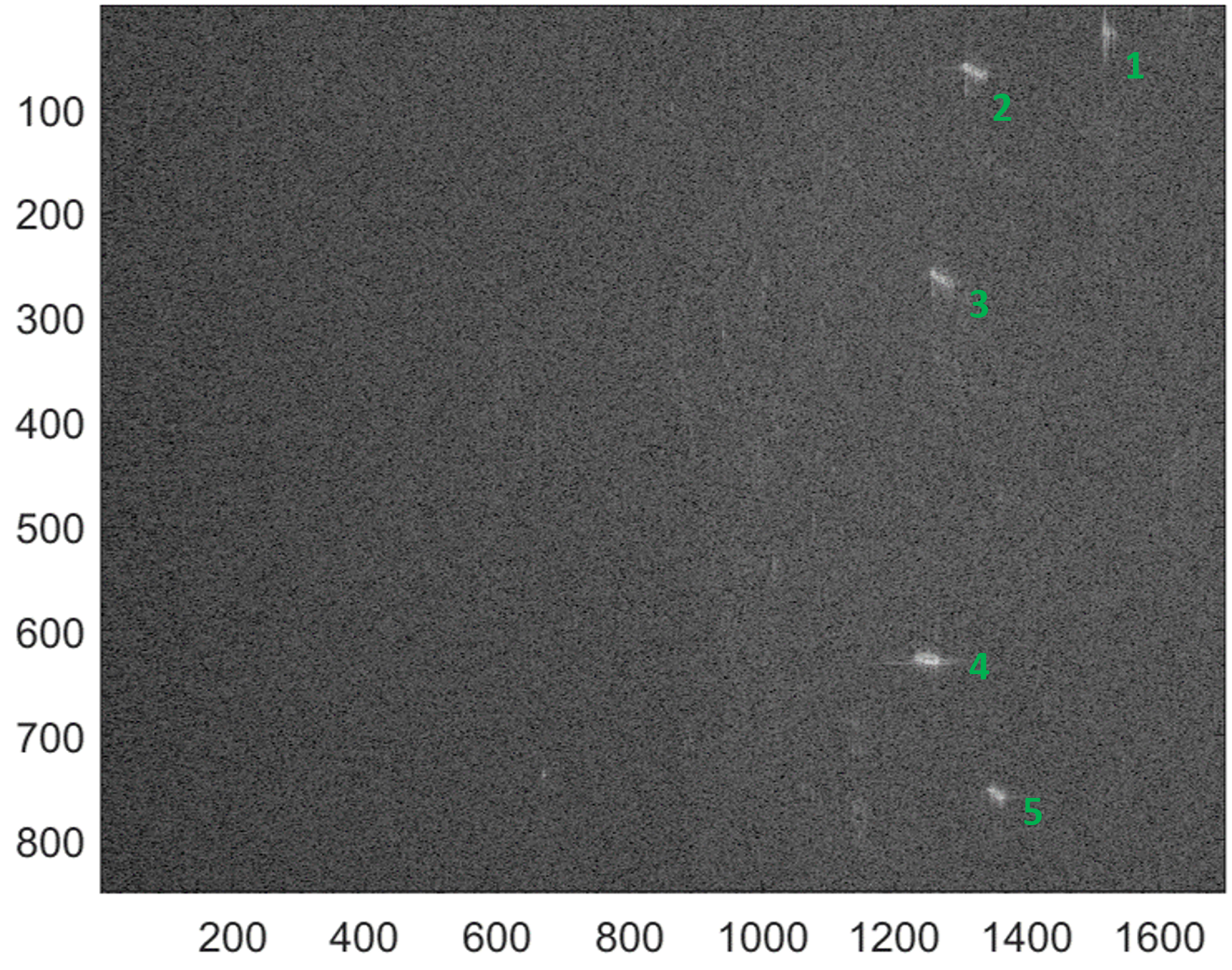}};
  \node[left=of img1, node distance=0cm, rotate = 90, xshift=1.8cm, yshift=-0.9cm,font=\color{black}] {{Along-Track [Samples]}};
  \node[below=of img1, node distance=0cm, xshift=0cm, yshift=1.1cm,font=\color{black}] {{Slant-Range [Samples]}};
  \node[below=of img1, node distance=0cm, xshift=0cm, yshift=0.6cm,font=\color{black}] {{(a)}};
\hspace{4cm}
  \node (img1)  {\includegraphics[height=3cm,width=3.5cm]{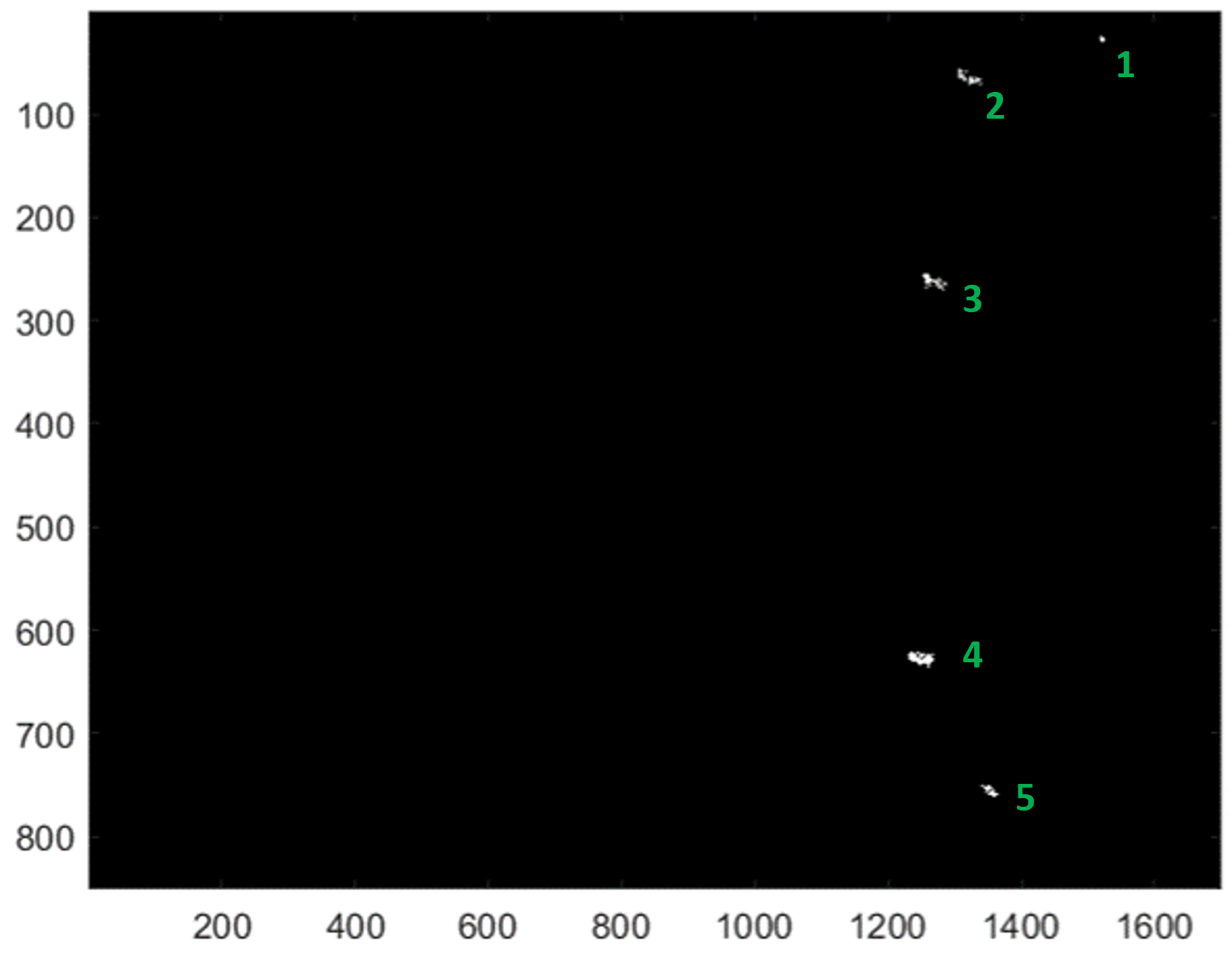}};
  \node[left=of img1, node distance=0cm, rotate = 90, xshift=1.8cm, yshift=-0.9cm,font=\color{black}] {{Along-Track [Samples]}};
  \node[below=of img1, node distance=0cm, xshift=0cm, yshift=1.1cm,font=\color{black}] {{Slant-Range [Samples]}};
  \node[below=of img1, node distance=0cm, xshift=0cm, yshift=0.6cm,font=\color{black}] {{(b)}};
\end{tikzpicture}
\caption{The result of applying the 2D CFAR algorithm to the part of the data-set shown as the area inside the green rectangle in Fig.~\ref{fig:EnglishBay_Speckle}. a) the ships in sea water, b) the detected ships.}
\label{fig:cfar_ships}
\end{figure}
\vspace{1 pt}
To achieve the result which has been displayed in Fig.~\ref{fig:cfar_ships}-(b), we have set the number of guard cells for the upper and lower wings to 60 and for the left and right wings to 90. We have also selected the number of training cells for the upper, lower, left, and right wings to be equal to 5. The probability of false alarm has been selected as $p_{\rm fa} = 10^{-6}$. The Weibull parameters are estimated as $\alpha = 1.9521$ and $\beta = 0.4835$. Consequently, based on (\ref{thr_weibull}), the adaptive threshold is calculated as $T_{\rm aW} = 1.856$. Moreover, the sample mean and standard deviation for the background clutter, which have been described in (\ref{detection}), are estimated as $\mu_c = 0.4286$ and $\sigma_c =  0.2294$.
We consider another case in which one of the ships, as has been highlighted as the area inside the yellow rectangle  in Fig.~\ref{fig:EnglishBay_Speckle}, is buried in the sea clutter. Moreover, it should be noted that, the image presented in Fig.~\ref{fig:EnglishBay_Speckle} is contaminated with the speckle noise.  
In Fig.~\ref{fig:img_ship}, we have presented the reconstructed image from a single ship which has been highlighted as the area inside the yellow box in Fig.~\ref{fig:EnglishBay_Speckle}. 
\begin{figure}
\begin{tikzpicture}[yshift=0.00001cm][font=\small]
  \node (img1)  {\includegraphics[height=3cm,width=3.5cm]{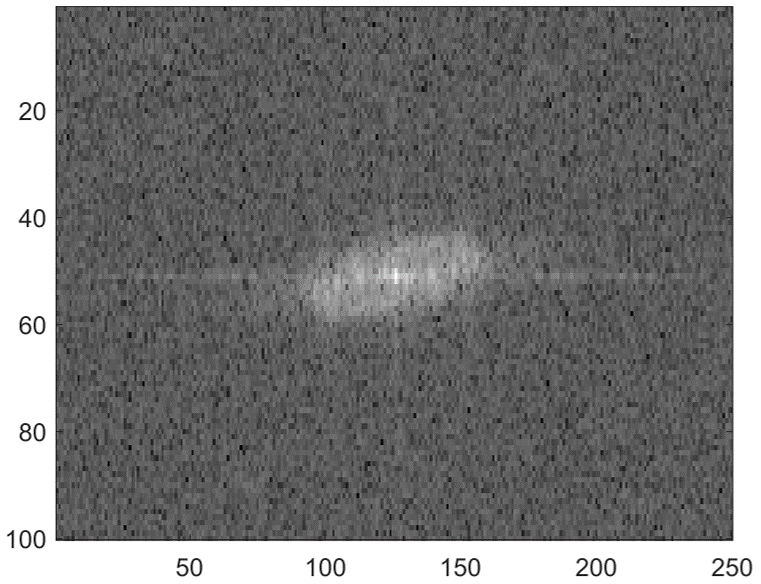}};
  \node[left=of img1, node distance=0cm, rotate = 90, xshift=1.8cm, yshift=-0.9cm,font=\color{black}] {{Along-Track [Samples]}};
  \node[below=of img1, node distance=0cm, xshift=0cm, yshift=1.1cm,font=\color{black}] {{Slant-Range [Samples]}};
  \node[below=of img1, node distance=0cm, xshift=0cm, yshift=0.6cm,font=\color{black}] {{(a)}};
\hspace{4cm}
  \node (img1)  {\includegraphics[height=3cm,width=3.5cm]{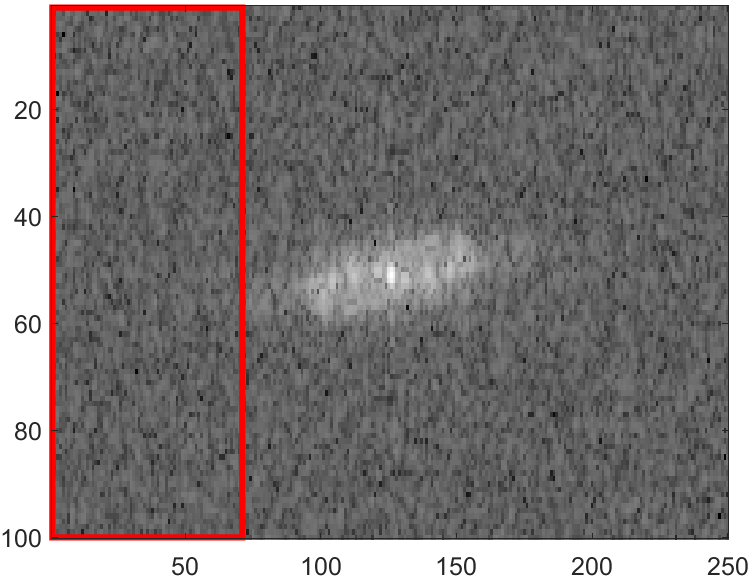}};
  \node[left=of img1, node distance=0cm, rotate = 90, xshift=1.8cm, yshift=-0.9cm,font=\color{black}] {{Along-Track [Samples]}};
  \node[below=of img1, node distance=0cm, xshift=0cm, yshift=1.1cm,font=\color{black}] {{Slant-Range [Samples]}};
  \node[below=of img1, node distance=0cm, xshift=0cm, yshift=0.6cm,font=\color{black}] {{(b)}};
\end{tikzpicture}
\caption{The reconstructed image for a single ship shown as the area inside the yellow rectangle in Fig.~\ref{fig:EnglishBay_Speckle}, a) no windowing, b) by applying Hamming window in both range and along-track directions to reduce the side-lobe levels.}
\label{fig:img_ship}
\end{figure} 
The part of the image, shown in Fig.~\ref{fig:img_ship}-(b), which has been highlighted as the area inside the red rectangle is utilized to estimate the histogram of the sea clutter and the result has been illustrated in Fig.~\ref{fig:hist_ship}-(a). 
\begin{figure}
\begin{tikzpicture}[yshift=0.00001cm][font=\small]
  \node (img1)  {\includegraphics[height=3cm,width=3.5cm]{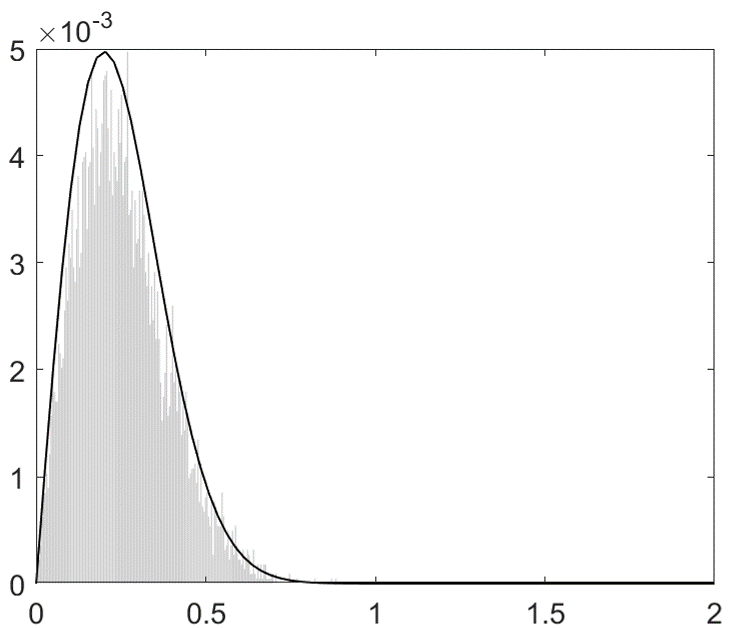}};
  \node[left=of img1, node distance=0cm, rotate = 90, xshift=1cm, yshift=-0.9cm,font=\color{black}] {{Frequency}};
  \node[below=of img1, node distance=0cm, xshift=0cm, yshift=1.2cm,font=\color{black}] {{$\sqrt{{I}^2 + {Q}^2}$}};
\hspace{4 cm}
  \node (img1)  {\includegraphics[height=3cm,width=3.5cm]{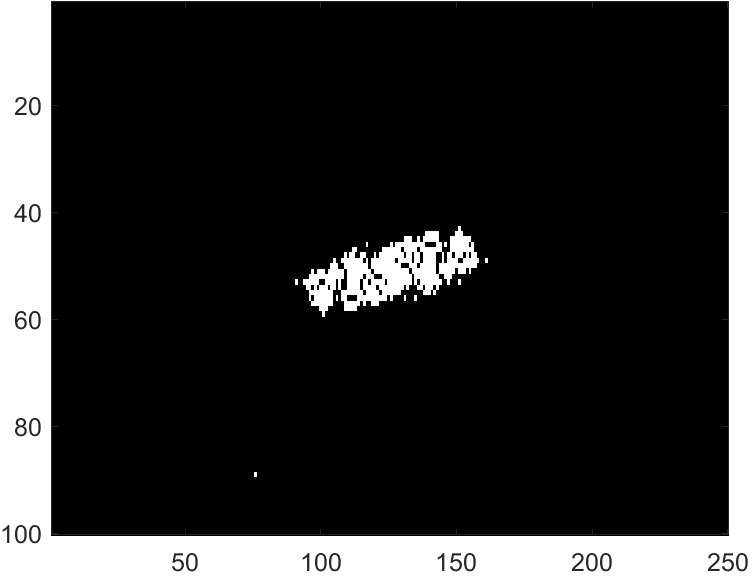}};
  \node[left=of img1, node distance=0cm, rotate = 90, xshift=1.75cm, yshift=-0.9cm,font=\color{black}] {{Along-Track [Samples]}};
  \node[below=of img1, node distance=0cm, xshift=0cm, yshift=1.1cm,font=\color{black}] {{Slant-Range [Samples]}};
\end{tikzpicture}
\caption{a) the histogram of the part of the reconstructed image shown in Fig.~\ref{fig:img_ship}-(b) inside the red rectangle as a representation of the statistical properties of the sea clutter. The estimated probability density function is based on the Weibull Distribution, b) the result of applying the 2D CFAR method to the image of the ship in sea clutter which has been illustrated in Fig.~\ref{fig:img_ship}-(b)}
\label{fig:hist_ship}
\end{figure} 
\vspace{1 pt}
As it is clear from Fig.~\ref{fig:hist_ship}-(a), the Weibull distribution can model the histogram with high accuracy. 
Next, we apply the 2D CFAR technique to the reconstructed image presented in Fig.~\ref{fig:img_ship}-(b). We have presented this remarkable result in Fig.~\ref{fig:hist_ship}-(b). To achieve the result presented in Fig.~\ref{fig:hist_ship}-(b), we have set the number of guard cells for the upper and lower wings to 35 and for the left and right wings to 60. We have also selected the number of training cells for the upper, lower, left, and right wings to be equal to 8. The probability of false alarm has been set to $p_{\rm fa} = 10^{-6}$. The Weibull parameters are estimated as $\alpha = 1.9912$ and $\beta = 0.2841$. Consequently, based on (\ref{thr_weibull}), the adaptive threshold is calculated as $T_{\rm aW} = 1.0621$. Furthermore, the sample mean and standard deviation for the background clutter, which have been utilized in (\ref{detection}), are estimated as $\mu_c = 0.2518$ and $\sigma_c =  0.1321$, respectively.

\section{Conclusion}
Target detection in sea clutter for SAR images was addressed. The statistical properties of the sea clutter was investigated thoroughly which based on that a 2D adaptive threshold was developed for target detection in sea clutter. Several relevant distribution functions were studied and the correct model based on the Weibull distribution was applied to the sea clutter. The KL divergence metric was utilized as the goodness-of-fit test. The validity of the proposed model was verified based on the experimental data and the results were presented.

%

%

\begin{IEEEbiography}[{\includegraphics[width=1in,height=1.25in,clip,keepaspectratio]{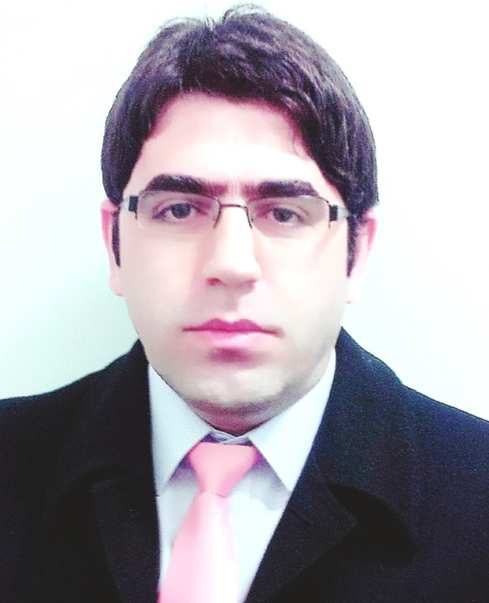}}]{Shahrokh Hamidi} was born in 1983, in Iran. He received his B.Sc., M.Sc., and Ph.D. degrees all in Electrical and Computer Engineering. He is with the Faculty of Electrical and Computer Engineering at the University of Waterloo, Waterloo, Ontario, Canada. His current research areas include statistical signal processing, mmWave imaging, Terahertz imaging, image processing, system design,  multi-target tracking, wireless communication, machine learning, optimization, and array processing.
\end{IEEEbiography}

\end{document}